\documentclass[useAMS,usenatbib]{mn2e}
\usepackage{graphicx}

\title[Phase-lags of solar hemispheric cycles]{Phase lags of solar hemispheric cycles}
\author[J. Murak\"ozy and A. Ludm\'any]{J. Murak\"ozy\thanks{E-mail:
murakozyj@puma.unideb.hu (JM)} and A. Ludm\'any\thanks{E-mail: ludmany@tigris.unideb.hu (AL)}\\
Heliophysical Observatory, H-4010 Debrecen P.O.B. 30. Hungary}

\begin{document}


\pagerange{\pageref{firstpage}--\pageref{lastpage}} \pubyear{}

\maketitle

\label{firstpage}

\begin{abstract}

The North-South asymmetry of solar activity is variable in time and strength. We analyse
the long-term variation of the phase lags of hemispheric cycles and check a conjectured
relationship between these phase lags and the hemispheric cycle strengths. Sunspot data 
are used from cycles 12-23 in which the separation of northern and southern hemispheres
is possible. The centers of mass of the hemispheric cycle profiles were used 
to study the phase relations and relative strengths of the hemispheric cycles. This 
approach considers a cycle as a whole and disregards the short-term fluctuations of 
the cycle time profile. The phase of the hemispheric cycles shows an  alternating 
variation: the northern cycle leads in 4 cycles and follows in 4 cycles. No significant 
relationship is found between the phase and strength differences of the hemispheric cycles.
 The period of 4+4 cycles appears to be close to the Gleissberg cycle and may provide 
a key to its physical background. It may raise a new aspect in the solar dynamo mechanism 
because it needs a very long memory.

\end{abstract}


\section{Introduction}

Asymmetry is a basic feature of astrophysical dynamos. All asymmetric properties and their temporal variations may contribute to the realistic dynamo modeling. A stationary magnetic field with perfect axial symmetry cannot be maintained by dynamo action according to the Cowling theorem but additional helical non-axisimmetric flow components can contribute to the maintenance of the field which is necessarily variable and asymmetric. This important feature has been investigated theoretically by \citet{Moss}, \citet{Tobias}, \citet{Gissinger}, \citet{Chatterjee} and \citet{Gallet}.

The solar north-south asymmetry and its variability were investigated in several works, most of them are based on sunspot data. \citet{Newton} did not find any regular variation, \citet{Carbonell} reported varying asymmetry, later  \citet{Ballester} detected a 43 year peak in the asymmetry periodogram confirmed by \citet{Zolotova07} by a different method. \citet{Chang} found periodicites between 9-12 years, \citet{Vizoso} reported a periodicity at 3.27 years and recently  \citet{Brajsa} reported a 70 years peak. \citet{Dajka} found asymmetry periods longer than 100 years by different methods. \citet{Lietal_02} and \citet{Lietal_09b} found a characteristic time scale of 12 Schwabe cycles more probable than a time scale of 8 cycles. The long-term variations may raise difficulties to the dynamo models which in the present form cannot give reliable forecasts even for two cycles ahead but fluctuations of about eight decades could hardly be accounted for in the frame of a dynamo mechanism.

Asymmetry properties were used by \citet{Javaraiah08} for an attempt of forecast for cycle 24.  \citet{Temmer02, Temmer06} also report varying asymmetry, being enhanced at cycle maximum. Asymmetries can also be studied by using different solar observables, \citet{Joshietal} studied active prominences, \citet{JoshiJoshi} analysed the soft X-ray index. \citet{Lietal_09a} found different cases for the connection between high and low latitude asymmetry. \citet{Mursula} detected a long-term variation in the N-S asymmetry of solar wind speed.

Extreme asymmetry may also appear temporarily. \citet{Sokoloff} reported a long period in the Maunder minimum when only the southern hemisphere was active in a narrow belt, they concluded that this was probably a mixed parity mode when a quadrupolar field became predominant instead of a dipole field. \citet{Bushby} pointed out that this situation may readily arise in a mean-field $\alpha \omega$ dynamo.

\begin{figure*}
\centering
\includegraphics*[angle=-90,width=15cm]{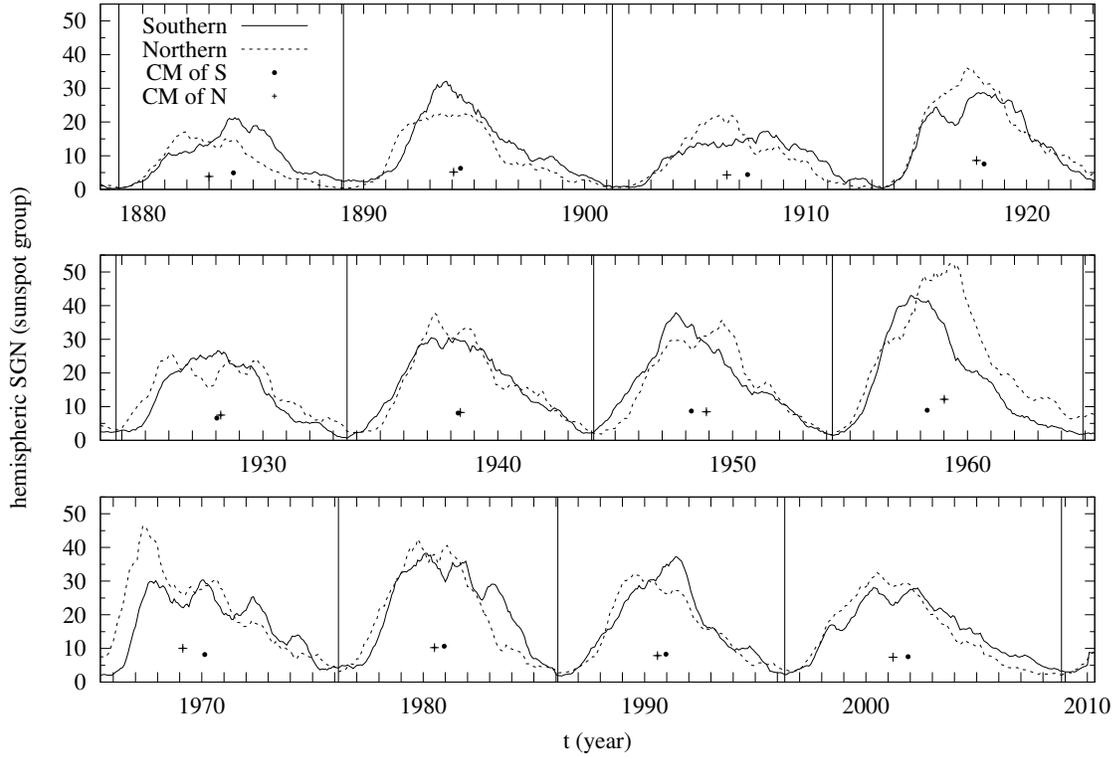}
 \caption{Profiles of cycles 12-23 smoothed with 11 month windows in the northern and southern hemispheres with the positions of their centers of mass. The vertical lines indicate the times of global minima. SGN means number of sunspot groups.}
	\label{centers}
\end{figure*}

\begin{figure*}
\centering
\includegraphics*[angle=-90,width=15cm]{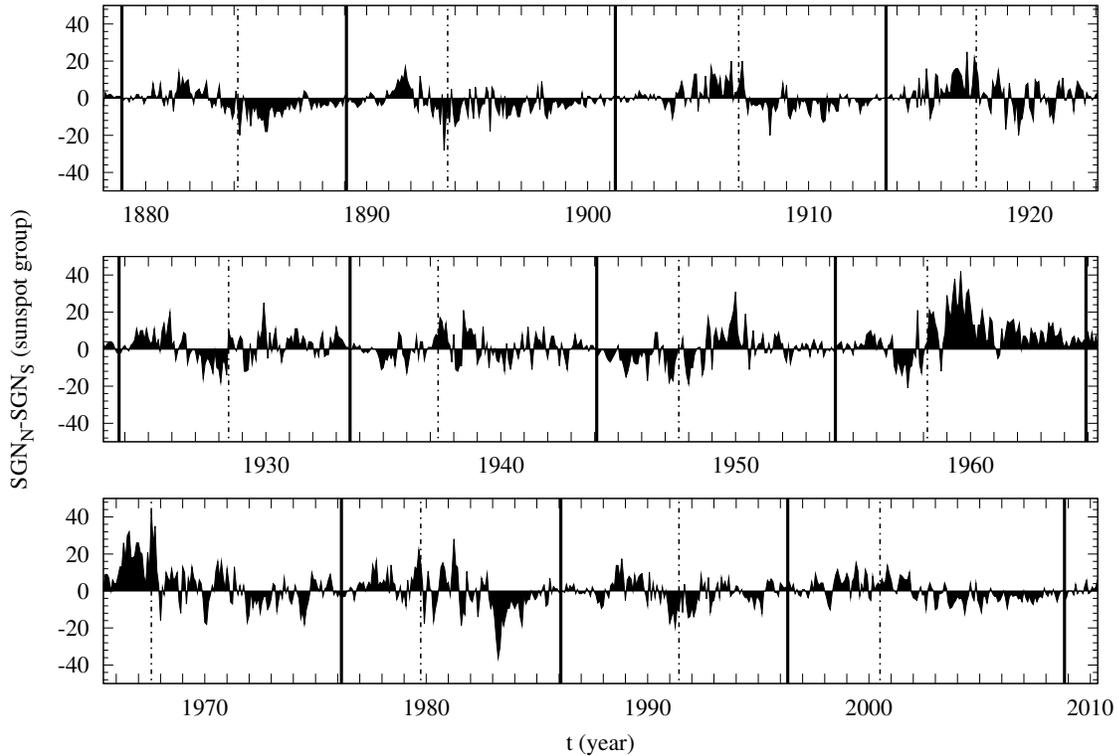}
 \caption{Monthly mean unsmoothed values of north-south activity differences in cycles 12-23. The cycles are separated with vertical continuous lines, the times of maxima are indicated with dashed lines.}
	\label{N-S_SGN}
\end{figure*}

We focus on the dynamics of the north-south asymmetry, in particular phase lags of hemispheric cycles and their possible connection to the relative strengths of northern and southern cycles.

\section{Sunspot data and methods}

Long-term features of solar north-south asymmetry can be studied by using sunspot catalogues containing sunspot positions. The catalogues provide data from 1874 to the present day. Our analysis is based on the Greenwich Photoheliographic Results \citep{GPR} for cycles 12-20, Kislovodsk sunspot group data \citep{Kislovodsk} for cycle 21 and Debrecen Photoheliographic Data or DPD \citep{Gyori} for cycles 22-23. The input data is the monthly mean number of sunspot groups, this is the only data which is continuously available since 1874 in the GPR and allows to distinguish between the hemispheres.

Systematic differences between different sources may distort the results, this is a common problem of all long-term analyses. The overlaps of the three sources were used for their intercalibration in two steps. In the first step the mean value of the monthly GPR/SD ratios were computed for the period 1966-1976, it is 1.172. The monthly SD values of the cycle 21 were multiplied by this factor, in such a way the SD dataset was calibrated to the GPR. In the second step the mean value of the monthly DPD/SD ratios (here the SD is the corrected dataset) were computed for the period 1986-1996, it is 1.380, then the monthly DPD values of the cycles 22 and 23 were divided by this factor. Thus the DPD was calibrated to the SD and, indirectly, to the GPR.

In order to determine phase and strength relations between hemispheric cycle profiles the strength and temporal position should be reliably established. However, the irregular shapes of cycle profiles, ambiguous maximum values, different ascending/descending slopes etc. make difficult this determination. Smoothing procedures with any arbitrary windows may input some subjective factors into the analysis, that we wanted to avoid. For this reason, the cycle profile is represented here by its center of mass. The y-coordinate of this point represents the strength of the cycle and the x-coordinate represents the date at which half of the sunspot groups have already appeared in that hemisphere in the given cycle. In this way the bulk of the cycle is considered regardless of its irregular shape, in other terms, the entire activity of each hemispheric toroidal field as a whole is taken into account. This approach disregards the occasional differences in phase and strength within the individual cycles which is not necessarily resulted by the same mechanisms as the possible long-term variation targeted in the present work.

Fig.~\ref{centers} shows the profiles of cycles 12-23 smoothed by 11 months running mean for the two hemispheres along with the centers of mass which have been computed from the unsmoothed, i.e. monthly mean sunspot group numbers (SGN). The center of mass of a cycle profile was computed in the following way. The  x-coordinate of the center (time) is the date at which the areas (the sums of monthly mean values, i.e. the Riemann integrals) of the  preceding and following half-profiles are equal. The y-coordinate of the center of mass point is the value at which the area above and below that point are equal for that hemisphere and that cycle. The times of minima separating the cycles are determined from the global activity time profile. Fig.~\ref{N-S_SGN} shows the north-south activity differences in monthly resolution. The hemispheric predominance has a considerable variability which is fairly stochastic and one cannot select the leading hemisphere on a shorter timescale than half a cycle. This makes difficult to determine the phase shifts, this is why the center-of-weight method is advantageous.

\begin{figure}
 \centering

\includegraphics[angle=-90,width=6cm]{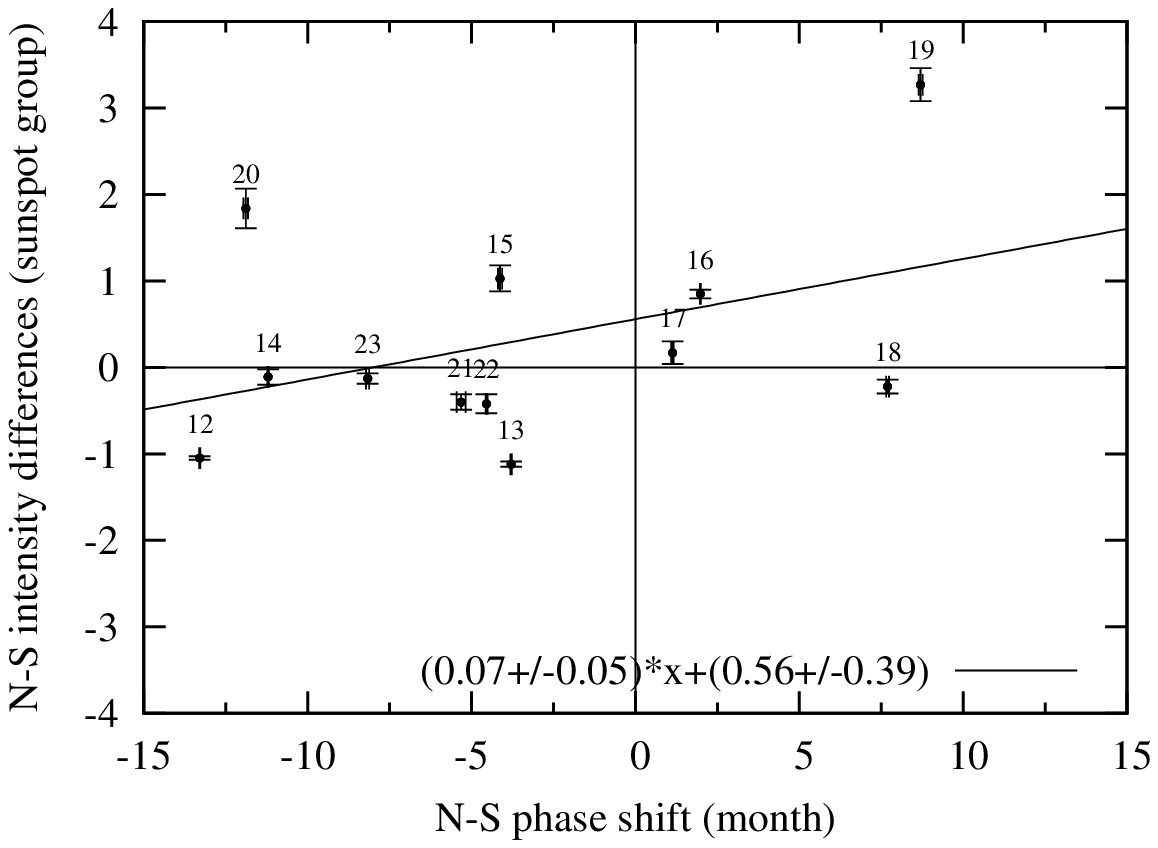}
\includegraphics[angle=-90,width=6cm]{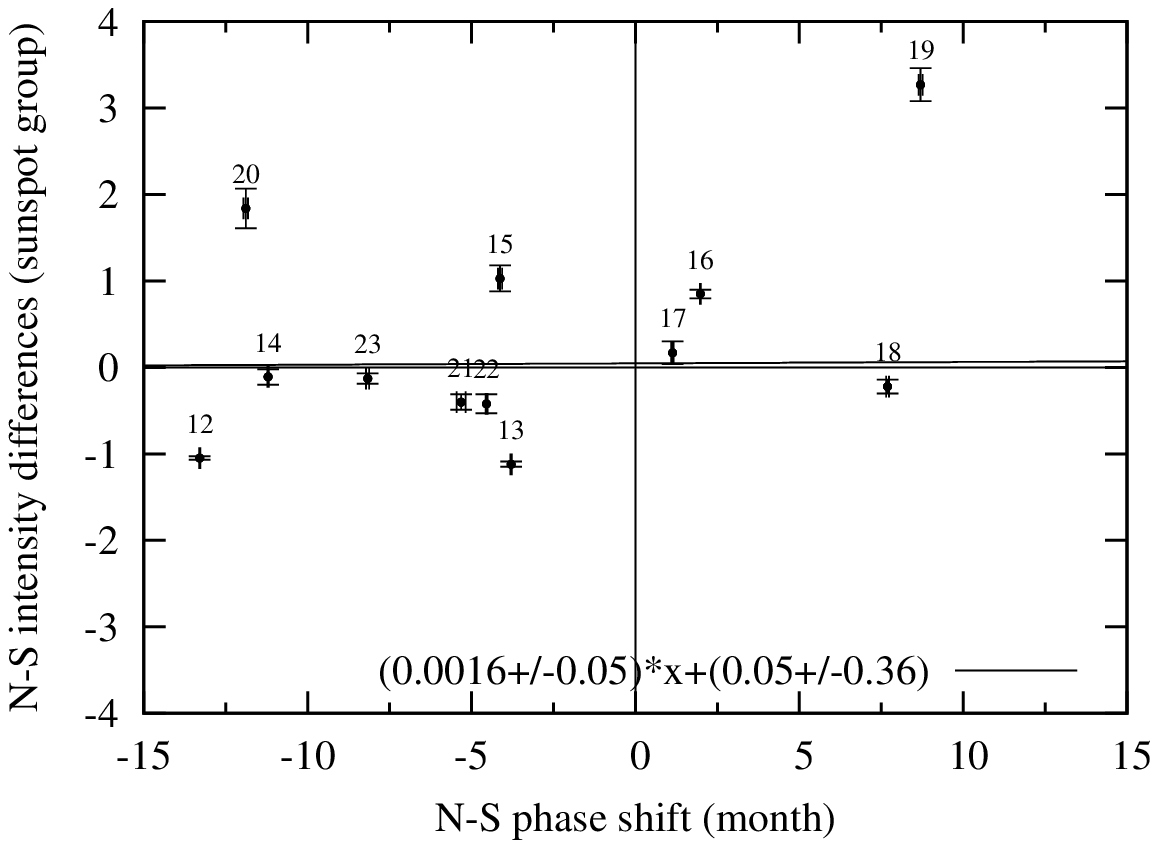}
 \caption{Upper panel: relation between the strength and phase differences of the centers of mass of hemispheric cycles. Lower panel: the regression line is computed by omitting cycle 19.}
	\label{NS_phaseintensity}
\end{figure}

The centers of mass provide a simple possibility to check whether the phase difference between the hemispheric cycles is related to their relative strength. The upper panel of Fig.~\ref{NS_phaseintensity} shows the comparison of the differences of intensities (y-axis) and dates (x-axis) of the centers of mass with fitted linear regression line. The southern values are subtracted from the northern ones for both quantities, so a negative x-value means leading northern cycle. The numbers of cycles are indicated at the points. Since the most deviating point belongs to cycle 19, the strongest cycle ever observed, the lower panel of the figure shows a regression line fitted by omitting this point. The weak trend completely disappears.

\begin{figure}
 \centering
\includegraphics[angle=-90,width=6cm]{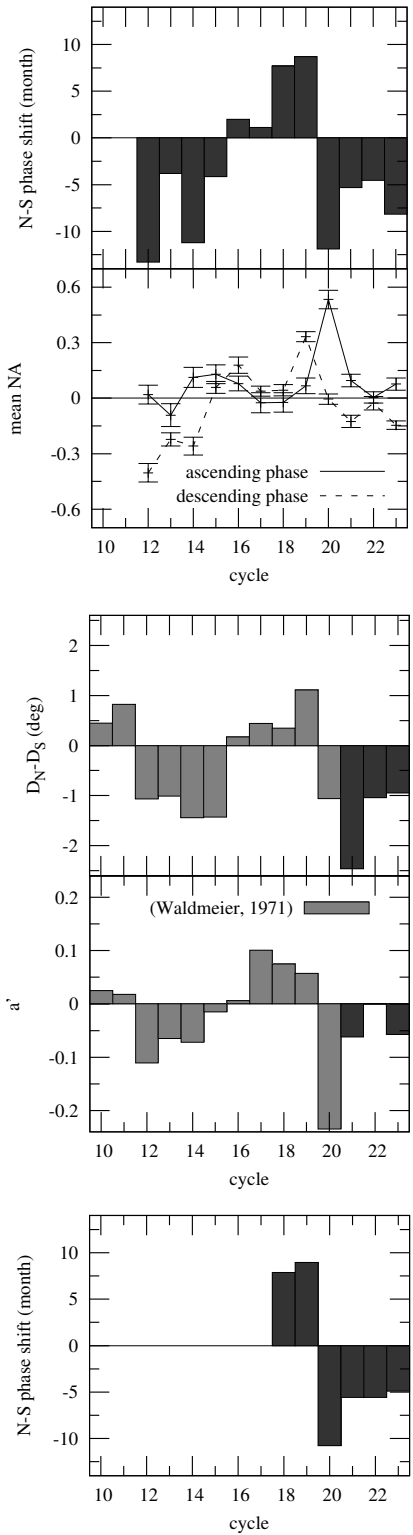}
 \caption{First panel: phases of the hemispheric cycles. Negative/positive bars mean leading/following northern cycles respectively. Second panel: variation of the north/south sunspot asymmetry index in cycles 12-23 for the ascending and descending phases separately. Third panel: hemispheric phase lags from the data of Waldmeier (1971) based on hemispheric Sp\"orer diagrams. Fourth panel: variation of the normalized asymmetry index (Waldmeier, 1971). Fifth panel: phase lags computed from the hemispheric sunspot numbers, \citet{Temmer06}.}
	\label{prec}
\end{figure}

The other investigated property is the long-term variation of the hemispheric cycle phase. By using the temporal differences of the centers of mass of cycle profiles the uppermost panel of Fig.~\ref{prec} shows the bar diagram of phases.  Similarly to the sign convention of Fig.~\ref{NS_phaseintensity}, negative values mean that the cycle of northern hemisphere is leading. 

This variation can also be studied by different methods. By using the $SGN_{N}$ and $SGN_{S}$ quantities, the monthly sums of sunspot groups for the northern and southern hemispheres where each group is counted once in a given month, the {\it NA} normalised asymmetry index between the northern/southern sunspot activity can be written in the following way:

\begin{equation}
      NA = \frac{SGN_{N} - SGN_{S}}{SGN_{N} + SGN_{S}}
\end{equation}

The asymmetry index can also be defined as the numerator of this formula but in this paper the above normalized form will be used. If the northern cycle profile is shifted ahead with respect to the southern one it means northern predominance in the ascending phase and southern predominance in the descending phase. Mean values of asymmetry index have been computed separately for the ascending and descending phases by averaging the monthly values in these periods, the result is plotted in the second panel of  Fig.~\ref{prec}. In contrast to Fig.\ref{NS_phaseintensity} and the first panel of Fig.\ref{prec}, in this case error bars can be rendered to the points because the monthly asymmetry values have a certain scatter in the ascending and descending periods. The error bars are smaller than the separations of the values in the two curves, so this variation can be regarded as real.

The calibration procedure mentioned in the second paragraph was not the only attempt to homogenize the dataset composed from three sources. Two other methods used the International Sunspot Number  \citep{SIDC}  as a normalization dataset to fit the datasets of the three catalogues to each other. The two procedures differed by the sampling periods. The results were practically the same in all cases. The greatest deviation from the magnitude values of Fig.~\ref{NS_phaseintensity} is 0.36 by the scaling of the figure. The greatest deviation from the time values of Fig.~\ref{prec} caused by the different methods is 0.68, by the scaling of the figure. This corresponds to about 20 days but most of the deviations are less than one day. This means that the homogenization procedure does not influence substantially the magnitude of these differential features and the variations plotted in the first two panels of Fig.~\ref{prec} are real.

The time span of the existing sunspot catalogues restricts the study of long-term variations. The only opportunity to extend the time span of the examinations is provided by the work of \citet{Waldmeier}. He used Zurich sunspot measurements for the investigation of hemispheric phase differences. This unpublished dataset covers the period of cycles 10-20. His methods are different from ours so they are suitable to check the reliability of the results. Waldmeier also published the numerical results in table VIII of his paper which are plotted here with grey bars in the third and fourth panels of Fig.~\ref{prec} to compare them directly with our diagrams. The north-south phase differences in the third panel of Fig.~\ref{prec} are determined by Waldmeier from the latitude differences of the hemispheric Sp\"orer diagrams which is a possible measure of the hemispheric phase difference. The closer is the mean latitude of activity to the equator the more advanced is the cycle. By Waldmeier's original designation the $D_{N}-D_{S}$ formula means the difference of northern-southern mean activity latitudes, the difference is averaged for an entire cycle. Negative value of $D_{N}-D_{S}$ means leading northern cycle. The diagram is completed with the cycles 21-23 with black bars, they are computed by using Waldmeier's procedure and the Kislovodsk-DPD datasets.

The time profile in the fourth panel of Fig.~\ref{prec} is obtained by Waldmeier from an asymmetry index analysis similar to our second procedure above (second panel of Fig.~\ref{prec}). Here the hemispheric sunspot data are properly equalized for the two hemispheres, a straight line is fitted to the yearly values of the asymmetry index and its steepness (marked by a') is taken as a measure of the hemispheric phase shift. The original figures of Waldmeier look differently, the present format is comparable with our plots. More detailed explanations are given in the original paper of \citet{Waldmeier}. This diagram is also completed with the cycles 21-23 by using Waldmeier's procedure and the Kislovodsk-DPD datasets.

A further different dataset is also suitable for checking the findings in a limited time span. \citet{Temmer06} published hemispheric sunspot numbers for the years 1945-2004. This means only six cycles but the time profiles can be compared at least in this interval. The bottom panel of Fig.~\ref{prec} shows the variation of hemispheric phase lags by using this dataset and the center of weight method (like the top panel). Cycle 23, the last one, is not complete in the dataset but the center of weight can be computed.

The present methods and those of \citet{Waldmeier} are of global nature in the sense that they consider the toruses globally disregarding the short-term intensity differences within the specific cycles as mentioned in the second section, i.e., each cycle gets a single phase lag value.  \citet{Li09} also used a global-type method, he applied cross-correlational analysis. The global approach cannot reveal short-term effects like the possible source of the Gnevyshev gap studied by \citet{Norton}, however, for long-term variations this may be more efficient.

\begin{figure}
 \centering
\includegraphics[angle=-90,width=8.8cm]{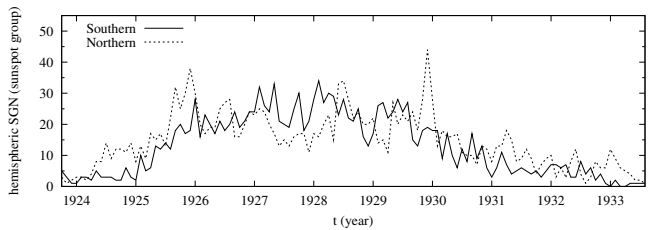}
 \caption{Monthly values of sunspot group numbers for both hemispheres in cycle 16.}
	\label{cycle16}
\end{figure}

\section{Results and discussion}

Fig.~\ref{NS_phaseintensity} does not support the conjecture that the hemispheric phases might be related to the relative strengths of the hemispheric cycles, no significant relation is found between the phase and strength differences. The apparent weak trend totally disappears by omitting the contribution of cycle 19 so the temporally leading role of a hemisphere does not mean its dominant role in intensity.

The hemispheric phase lags, however, show a characteristic variation. The panels of Fig.~\ref{prec} show fairly similar time profiles obtained by fairly different methods. The first one uses the time and intensity differences of the centers of weight of hemispheric cycle profiles. Its distinctive feature is the alternating leading-following-leading role of the northern hemisphere in 4-4-4 cycles. Specifically, the leading role of the northern hemisphere from cycle 12-15 is gradually exchanged to a leading southern hemisphere from cycle 16-19 and then, by an abrupt reversal, the northern hemisphere takes over the leading role again from cycles 20-23.

The rest of the panels help to check the reality of this variation. The asymmetry index variation in the ascending and descending phases (second panel) means the following. If the northern cycle is leading, as e.g. in cycles 12-15, then in the ascending phase the activity level of the northern hemisphere is higher than that of the southern hemisphere and the case is opposite in the descending phase, assuming that the lengths of both cycles are similar which is practically fulfilled in all cases. This means that the asymmetry index in the ascending phase is higher than that in the descending phase. The case is opposite in cycles 16-19 when the southern cycle leads. There is a similarity between the variations of the bar lengths of the first panel and the differences of descending-ascending values in the second one. It should be admitted that in this simple way the asymmetry index cannot be a measure of phase shifts between cycle profiles because it uses the data of activity level instead of time. However, the variation of the relative strengths of ascending/descending phases is in accordance with the variation of hemispheric phase lags.

It should be admitted that the time interval of 12 cycles is not long enough to esteem the long-term stability of this alternation. The reconstructed time-lag diagrams of \citet{Waldmeier} (third and fourth panels of Fig.~\ref{prec}) completed with the last three cycles corroborate the existence of the variation of 4+4 cycles in several ways. The observational material, the covered interval and the applied methods are different from ours but the detected temporal variation is basically the same. Cycle 22 has a weak but negative value in panel 4. The differences between the specific profiles can be attributed to the different methods, e.g., the center-of-weight method and the latitudinal difference (Sp\"orer) method use absolutely different indicators of the hemispheric phase difference. Nevertheless, a common feature is recognizable: after the four south-leading cycles an abrupt reversal initiates the next north-leading group of cycles. The most important additional information is that the southern hemisphere leads in the cycles 10-11 not covered by the Greenwich catalogue, so the effect does work on an extended interval too. The data of \citet{Li09} cannot be plotted here because he did not publish them numerically but his figures are in accordance with the present diagrams.

The last panel of Fig.~\ref{prec} is restricted in time but it is very useful for checking the variation. The hemispheric sunspot number (\citet{Temmer06}), its definition, determination and observational source, are independent from those applied by the procedures of the first four panels. The time profile was computed with the center-of-weight method. The phase lags unambiguously have the same variations in this restricted time span as in the rest of the panels.

Apparently the most ambiguous case is cycle 16 (Figs.\ref{centers} middle row, first cycle). The decaying phase of this cycle has northern predominance implying southern leading but exceptionally the rising phase also has a weak northern excess which might be the signature of northern leading. For this reason the approach of \citet{Zolotova09} is different from the mentioned global treatments. They follow the method of cross-recurrence plots and they assume that the variation of the phase lags may be independent from the cyclic cadence. They state that the N/S phase-lag changes sign at the maximum of the 16th cycle. 

This cycle is worth scrutinizing in detail, see the first cycles in the middle rows in Figs.\ref{centers} and \ref{N-S_SGN}. The north-south activity difference fluctuates strongly and it is obvious, as in all cycles, that this fluctuation does not imply the same fluctuation in the phase lags of hemispheric cycles in a short time frame (about one year). This correspondence cannot be made in a time frame shorter than a half cycle without any subjective decisions. The smoothing procedure with an arbitrary window is also a subjective contribution to the examination of time profile. This can be clearly seen in Fig.\ref{cycle16} where the monthly values are plotted for both hemispheres. The northern cycle profile is broader with two high peaks in 1925 and 1929 and it exhibits an unusually deep Gnevyshev gap. For the majority of the cycle the southern hemisphere seems to lead, except the first two years. However, by all mentioned global methods cycle 16 is lead by the southern hemisphere as can be seen in Fig.~\ref{prec}. The phase difference of the hemispheric activities characterizes the differing advanced states of the toruses over the entire cycle. This is the reason why the present work disregards the short-term fluctuations and follows global approaches.

\begin{figure}
 \centering
\includegraphics[angle=-90,width=6cm]{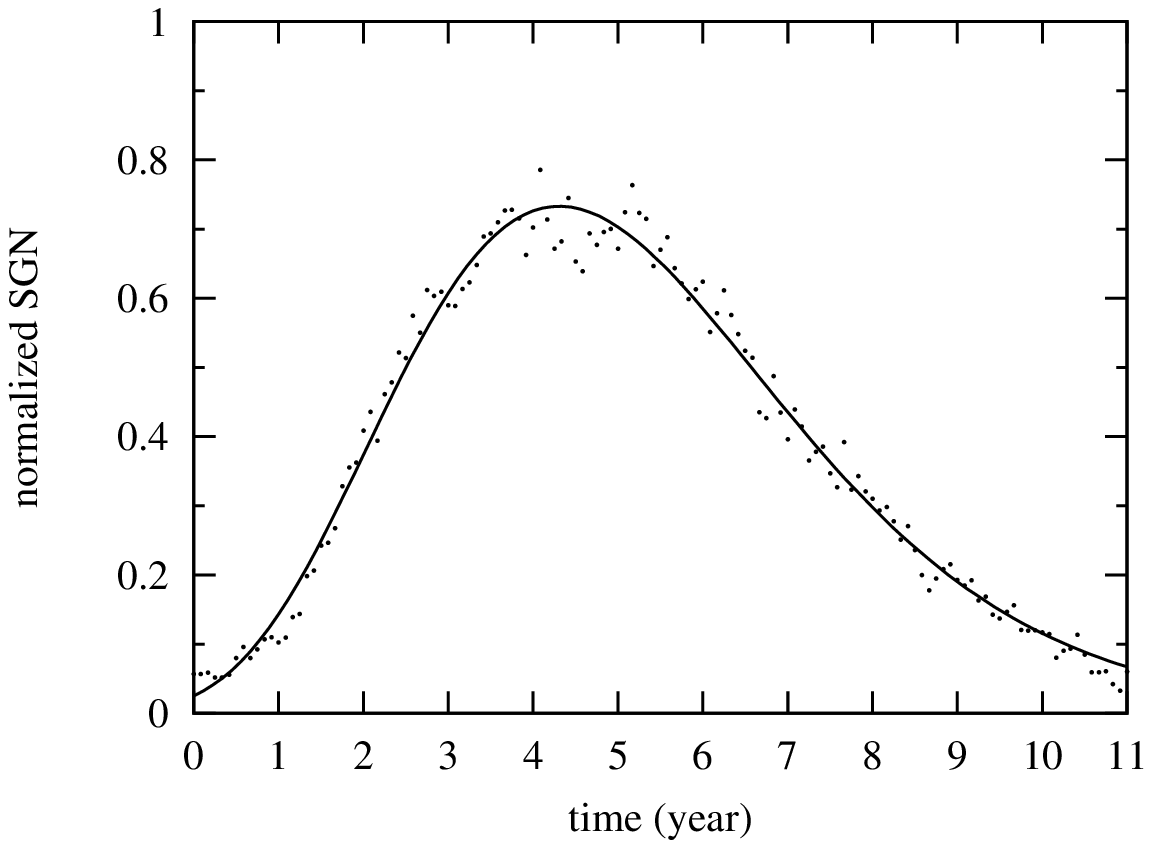}
\includegraphics[angle=-90,width=6cm]{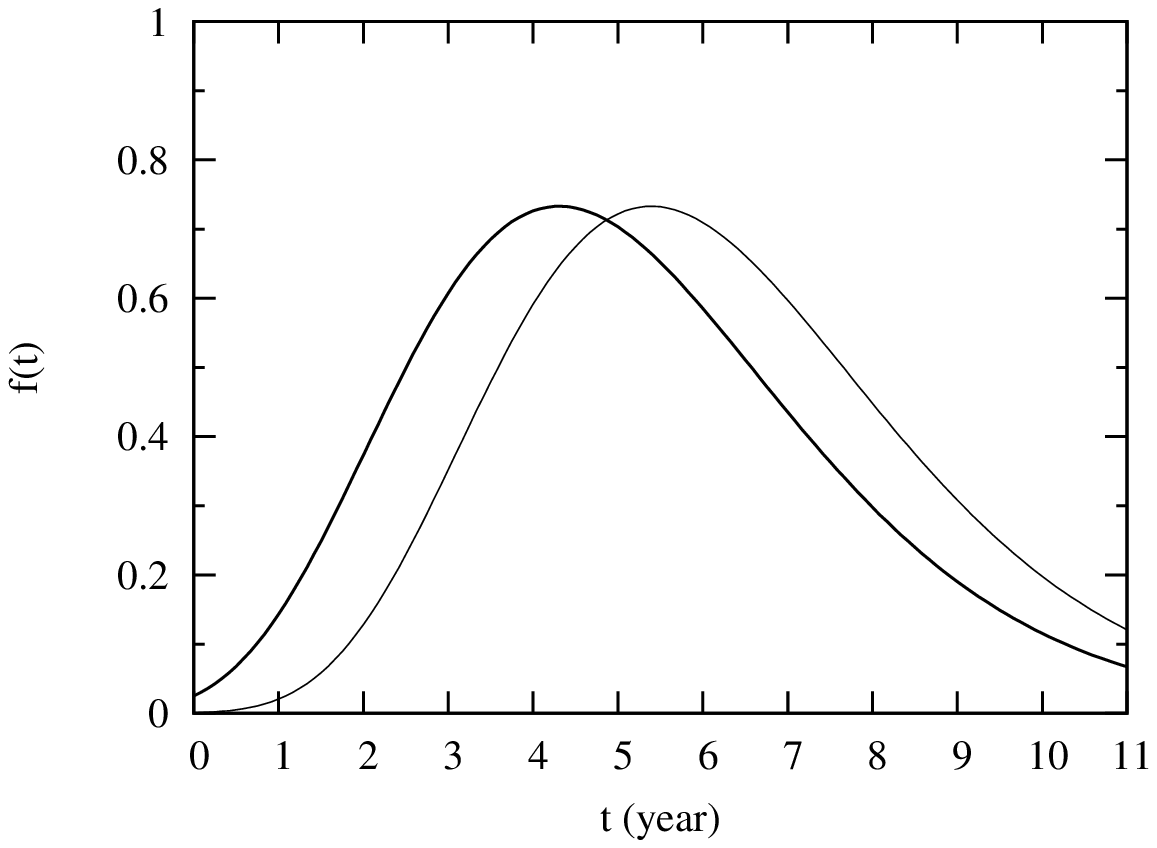}
\includegraphics[angle=-90,width=6cm]{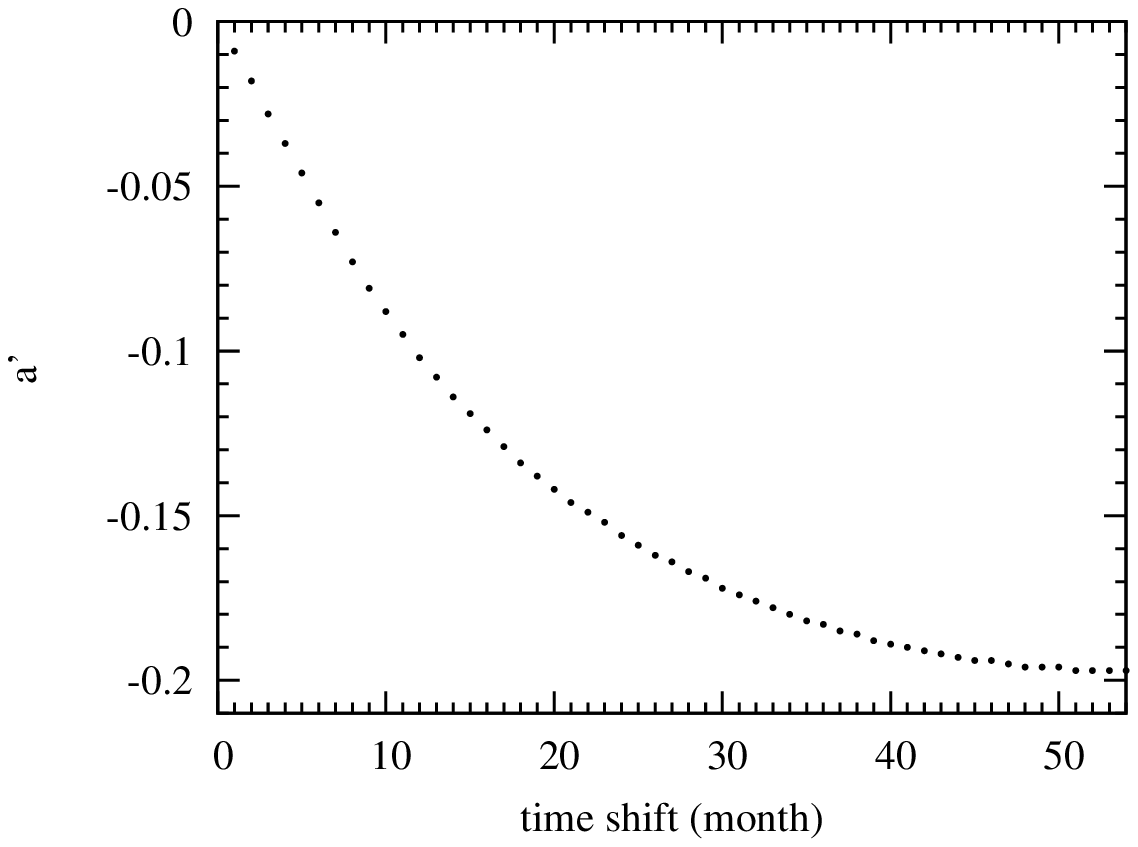}
 \caption{Top panel: mean time profile of cycles 12-23 and the fitted (2) curve. Second panel: identical N and S cycle profiles shifted by 13 months. Third panel: connection between the hemispheric phase lag and the steepness of asymmetry index variation (a') in the course of the cycle.}
	\label{calib}
\end{figure}

To assess the significance of the detected variation one should examine the relations between the different parameters used. Two methods refer to the same event: the mutual temporal shift of the time profiles of two cycles. The most unambiguous method is based on the centers of weight. However, the methods using asymmetry index in the ascending/descending phases (Fig.~\ref{prec}, panels 2 and 3) are only suitable in a restricted time span. The top panel of Fig.~\ref{calib} shows the mean time profile of cycles 12-23 computed with equalized cycle lengths and amplitudes. The following asymmetrical function has been fitted to the mean profile:

\begin{equation} 
      f(t)=H\cdot exp {(-\frac{(t-M)^2}{D(1+At)})} 
         \label{asymcurve} 
\end{equation}

In this formula {\it H} is the height of the curve, {\it M} is the position of maximum, {\it D} and {\it A} characterise the width and asymmetry of the curve. The second panel of Fig.~\ref{calib} shows two identical profiles, representing the northern and southern cycles, shifted by 13 months. The third panel shows how the steepness of asymmetry index variation during the cycle depends on the phase lag between the hemispheric cycles (Waldmeier's first method). The two parameters have nearly linear relationship at moderate phase lags but at higher shifts the curve is nonlinear. The center-of-weight method gives the most unambiguous measure of the phase shift, the asymmetry index method is only reliable in case of phase lags shorter than about 15 months. This is fulfilled in all cases.

The above two methods examine two different aspects of the same configuration: the shifted positions of two cycle profiles. The differing advanced states of the hemispheric cycles are also examined here by the differing rates of the equatorward motion of activity (Waldemeier's second method, third panel of Fig.4.), however, its relation cannot be examined mathematically with the methods of cycle profile shifts. These are two different manifestations of the cycle progress which can only be connected mathematically through arbitrarily chosen dynamo models. Therefore, the results of these two approaches can be regarded as independent checks of the same process.

\section{Conclusions, a long-term hemispheric wave}

Fig.~\ref{prec} presents a specific kind of long-term variation: alternating phase differences of hemispheric cycles with a cadence of 4+4 cycles. Similar results were obtained by \citet{Vizoso} and \citet{Li09}. \citet{Zolotova09} also found alternating phase lags of similar length but in their opinion this variation is not confined to the cyclic cadence. \citet{Javaraiah} also detected a 90-year period in the variation of the differential rotation B parameter, i.e. the latitudinal gradient of angular velocity.

There is no guarantee that the phase lag will always vary with integer multiples of four cycles. If the mechanism controlling this variation is not part of the solar dynamo then in the long term the cadence of 4+4 cycles may vary. By now, however, it is safe to say that this variation persisted during the last 14 cycles. The length of eight cycles nearly corresponds to the Gleissberg cycle \citep{Gleissberg}. 

The theoretical background is unclear. It might be a challenge to incorporate such a long period into self-consistent dynamo models because this would imply a process of very long memory spanning over several individual cycles. A further problem is that no connection has been found with other solar variations up to now, an example is the absence of phase-strength relationship in Fig.~\ref{NS_phaseintensity}. According to \citet{Norton} the phase lag and cycle length are not correlated either. Another relationship was reported by \citet{Waldmeier} between the phase lag and relative sunspot number over cycles 10-20 but it is no longer valid in cycles 21-23.

The phase lag is usually treated in terms of hemispheric coupling. Synchrony is more interesting theoretically, i.e. to understand the mechanism behind the hemispheric coupling which apparently supports the approximate symmetry of the Sp\"orer-diagram. \citet{Charbonneau} conducted model computations by applying weak hemispheric coupling through magnetic diffusion in the Babcock-Leighton (advection-dominate) model. He obtained quasi-periodic but not sign-changing variation of hemispheric lag by assuming stochastic forcing of the hemispheric dynamo numbers. By applying strong stochastic forcing he was also able to produce a Maunder-like grand minimum in both hemispheres simultaneously, but a single hemisphere grand minimum also happened in the computed time variation. Charbonneau concluded that the hemispheric coupling is certainly more effective than the recently conceived mechanism through magnetic diffusion and meridional flows might also be involved.

The apparent regularity of the hemispheric phase lag variations can hardly be interpreted by stochastic ingredients. As a possible contribution the impact exerted by the solar inertial motion can be considered. Following the papers of \citet{Jose} and \citet{Fairbridge} several works have been devoted to the apparent similarities between the long-term envelope of the solar cycles and the solar inertial motion  e.g. by \citet{Shirley06, Shirley09}, \citet{Wilson}, \citet{Charvatova} and \citet{Landscheidt}. The only theoretical work was published by \citet{Zaqar} for a possible mechanism driven by outer impact. \citet{deJager} published arguments against this kind of impact on the basis of an order-of-magnitude analysis. This means that the solar inertial motion cannot be the cause of the cyclic activity but perhaps its modulating effect cannot be excluded. \citet{Juckett} analysed connections of solar motions with the variations of N-S asymmetries but he did not report a period of eight-cycles. The question remains open.

\section*{Acknowledgments}

The research leading to these results has received funding from the European Community's Seventh Framework Programme (FP7/2007-2013) under grant agreement n° 218816. The authors are grateful for the referee's comments and suggestions which significantly improved the manuscript.

\end{document}